\documentclass[prl,twocolumn,nofootinbib,aps,superscriptaddress,showpacs]{revtex4-1}
\usepackage{color,graphicx,shortvrb,epsfig}

\usepackage{graphicx}
\usepackage{dcolumn}
\usepackage{bm}

\usepackage{colordvi}
\usepackage{psfig,epsfig,amssymb,amsmath,latexsym}

\newcommand{\no}{\nonumber}
\def\lsim{\mathrel{\rlap{\lower4pt\hbox{\hskip1pt$\sim$}}
    \raise1pt\hbox{$<$}}}         
\def\gsim{\mathrel{\rlap{\lower4pt\hbox{\hskip1pt$\sim$}}
    \raise1pt\hbox{$>$}}}         

\def\lsim{\mathrel{\rlap{\lower4pt\hbox{\hskip1pt$\sim$}}
    \raise1pt\hbox{$<$}}}         
\def\gsim{\mathrel{\rlap{\lower4pt\hbox{\hskip1pt$\sim$}}
    \raise1pt\hbox{$>$}}}         

\def\beq{\begin{equation}}
\def\eeq{\end{equation}}
\def\ba{\begin{eqnarray}}
\def\ea{\end{eqnarray}}

\def\<{\langle}
\def\>{\rangle}
\def\ra{\rightarrow}

\begin{document}

\begin{flushright}
\date{}
\end{flushright}

\title{Bending elasticity of macromolecules: \\
analytic predictions from the wormlike chain model
}
\author{Anirban Polley}
\author{Joseph Samuel} 
\author{Supurna Sinha}
\affiliation{Raman Research Institute, Bangalore, India 560 080}


\begin{abstract}
We present a study of the bend angle distribution of semiflexible polymers
of short and intermediate lengths within the wormlike chain  model.
This enables us to calculate the elastic response of a stiff molecule to
a bending moment. Our results go beyond the Hookean regime and explore the
nonlinear elastic behaviour of a single molecule. We present analytical 
formulae for the bend angle distribution and for the moment-angle
relation. Our analytical study is compared against numerical Monte Carlo 
simulations. 
The functional forms derived here can be applied to fluorescence microscopic
studies on actin and DNA. Our results are relevant to recent studies
in ``kinks'' and cyclization in short and intermediate length DNA strands.
\end{abstract}

\maketitle

\section{Introduction}
\label{intro}
A classic study in elasticity is the bending of beams or rods subject to forces
and moments\cite{landau}. At the cellular level there are many macromolecular 
structures like actin
and cytoskeletal filaments, which are like beams in giving rigidity 
and structure to the cell. Unlike the beams studied by civil engineers, 
these macromolecular beams are subject to thermal fluctuations. 
The purpose of this study is to look at the role of thermal fluctuations
in shaping the elastic properties of macromolecular beams.

We work within the wormlike chain model\cite{kratkyporod,doi}, 
which has been known to describe
double stranded DNA \cite{siggia} as well as actin 
filaments\cite{wilhelmfrey}. 
For clarity we consider an experiment
in which one end of the molecule is fixed at the origin and its tangent
vector at the same end is constrained to lie 
along the $\hat{z}$ direction. We wish to know the number
of configurations (counted with Boltzmann weight) that will  result
in the final tangent vector $\hat{t}_f$. 
In this paper we present an approximate analytical study of this statistical
mechanical problem.

Some earlier treatments of this problem\cite{stiff2,stiff1} restrict to small
bending angles, so that the polymer is essentially straight. One can then
replace the sphere of tangent directions by the tangent plane to this sphere.
This gives a good account of small bending angles. 
However, there is considerable experimental
and theoretical interest in large bending angles 
to understand cyclization
of DNA \cite{cloutier,mazur,Nelson}. Ref. 
\cite{fluct} developed a new approximation technique that works even for 
large bending angles, say of order $\pi/2$.  The treatment of \cite{fluct}
is general and includes applied forces and torques. In this paper
we apply the general theory to a special case to illustrate its use:
we treat the pure bending elasticity of a semiflexible polymer not
subject to stretching forces or twisting torques.

Two different experimental techniques are possible to probe the elastic
properties. 

{\it Distribution of tangent vectors:}
One can tag the 
ends of the molecule with flourescent dye\cite{legoff} 
to determine the direction of the initial and final tangent
vectors. By fluorescence video microscopy, one finds the angular 
distribution $P(\theta)$ of the bending angle $\theta$ , where
$\theta$ is defined by $\cos{\theta}=\hat{t}_i.\hat{t}_f$.  
This experimental technique has been used to study actin in 
two dimensional studies in Ref.\cite{legoff}. 
The angular distribution of $\theta$ gives us the elastic properties of
the molecule and one can compute, for instance, the probability for a given bending
angle $\theta$ and compare with the theoretical expectation.

{\it Measuring moment angle relations:} 
A more invasive experimental technique is to tether one end of the molecule,
attach a magnetic bead to the other and apply bending moments to the molecule
by varying the direction of an applied magnetic field. 
Note that we do not constrain the final position of the molecule $x(L)$ 
but only its final tangent
vector $\hat{t}_f$.
A uniform magnetic 
field will result in a pure bending moment without applying any stretching 
force.
Plotting the bending moment vs the bending angle gives another experimental
probe of the elastic properties. In this paper we derive the predictions of
the wormlike chain for both these experimental situations.

We first derive an approximate analytical formula for the free energy 
allowing for thermal fluctuations of a wormlike chain polymer. We then
plot the expected theoretical distribution of bending angles and
the theoretically expected moment-angle relation.
We conclude with a discussion.

\section{Mechanics and Fluctuations}
\label{mechanicsandfluctuations}

In the simplest wormlike chain model,
we model the polymer by a space curve $\vec{x}(s)$. 
 $\vec{x}(s)$ describes 
the  curve, and $\hat{t}(s)  =  \frac{d\vec{x}}{ds}$, its tangent vector. 
$s$ is the arc length parameter along the curve ranging from $0$ to $L$, the 
contour length of the curve. 
${\vec x}(0)=0$ since one end 
is fixed at the origin. The tangent 
vectors at both ends ${\hat t}(0)$,${\hat t}(L)$ are fixed to $\hat{t}_i$ and
$\hat{t}_f$ respectively. 

The mathematical problem we face is to compute the partition function
\begin{equation}
Q(\hat{t}_i,\hat{t}_f)= \sum_{{\cal C}} \exp - \Big[\frac{{\cal E}({\cal 
C})}{k_B T}\Big]\;. 
\label{ptnfull}
\end{equation}
In Eq.(\ref{ptnfull}), the sum is over all allowed configurations of the 
polymer, those which satisfy the boundary conditions
for the tangent vector at the two ends: 
$\hat{t}_i=\hat{t}(0)=\hat{z},\hat{t}_f=\hat{t}(L)$. 
The energy functional is given by
\begin{equation}
{\cal E} ({\cal C}) = \frac{A}{2} \int^{L}_{0} 
(\frac{d\hat{t}}{ds}.\frac{d\hat{t}}{ds}) ds,
\label{energy}
\end{equation}
where $A$ is an  elastic constant with dimensions 
of energy times length. The quantity $L_p=A/kT$ is 
the persistence length. For example actin has a persistence
length of about $16 \mu m$ and an elastic constant of $6.7\times 10^{-27}$
in units of Nm${}^2$. Lengths of actin filaments can go up to 
hundreds of microns, so the molecule is of intermediate flexibility.

We will compute the partition function assuming that near the stiff limit 
($L$ not much larger than $L_p$), 
the sum over curves is dominated by configurations near the 
minimum of the energy.
\begin{figure}[h!t]
\includegraphics[width=8.0cm]{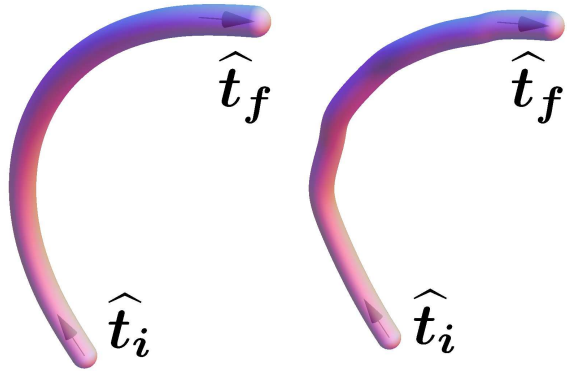}
\caption{Two configurations of a molecule with  
fixed end tangent vectors $\hat{t}_i$ and $\hat{t}_f$. 
(left) Minimum energy configuration and (right)
the same configuration is shown slightly perturbed by thermal fluctuations.
}
\label{cartoon}
\end{figure}

The minimum energy configurations of the polymer are those where the tangent
vector $\hat{t}(s)$ describes the shortest geodesic connecting $\hat{t}_i$
to $\hat{t}_f$ (Fig. 1).
The minimum energy is easily computed to be 
\begin{equation}
E_{min}=A \theta^2/(2L),
\label{minimum}
\end{equation}
which is  a purely mechanical contribution to the energy\cite{landau,mazur}.

To go beyond mechanics and include the effect of thermal fluctuations (Fig. 1),
we use an approximation described in \cite{fluct}.
The approximation consists of using an expansion of the 
energy about the minimum  energy configuration and 
retaining fluctuation terms about the minimum 
to quadratic order. This calculation was done in general form 
in  Eq.(53) of Ref. \cite{fluct} 
for a single molecule subject to forces and twisting torques. 
The determinant of the fluctuation operator ${\cal O}$ is calculated
in \cite{fluct}. In the present case, this fluctuation 
determinant can be calculated 
either by specialising  Eq.(53) of \cite{fluct} to the case of zero force and torque
of \cite{fluct} or by the method described in \cite{epl}
in the context of Brownian motion on the sphere. As is to be expected,
both methods give the same answer
\begin{equation}
Det{\cal O}= L^2 \sin{\theta}/\theta.
\label{determ}
\end{equation}
Performing the Gaussian integration over the fluctuations we find
the distribution function 
$1/\sqrt{(Det {\cal O})}\exp{-\frac{E_{min}}{kT}}$
and we arrive at the approximate
formula
\begin{equation}
Q(\theta)=\frac{{\cal N}(L)}{L} \sqrt{\frac{\theta} 
{\sin{\theta}} }\exp{[-\frac{A\theta^2}{2LkT}]}
\label{Qoftheta}
\end{equation}
for the partition function as a function of the final angle $\theta$,
where ${\cal N}$ is a normalization constant to be determined by
the normalization condition 
\begin{equation}
\int_0^\pi Q(\theta) \sin{\theta} d\theta=1
\label{norm}
\end{equation}
which includes the measure $\sin{\theta}$ on the sphere.

The result (\ref{Qoftheta}) implies that the number of configurations
with a final tangent vector ${\hat t}_f$ making an angle
$\theta$  defined 
by $\arccos{(\hat{t}_i.\hat{t}_f)}$ is given by 
$P(\theta)=Q(\theta)\sin{\theta}$ leading to the 
simple analytic approximate formula for the bend angle
distribution
\begin{equation}
P(\theta)=\frac{{\cal N}}{L} \sqrt{\theta\sin{\theta}}
\exp{{[-\frac{A\theta^2}{2LkT}]}}
\label{Poftheta}
\end{equation}
for $\theta$. Note that this is a closed analytic form rather than 
an infinite series\cite{epl}. It is thus suitable and convenient for experimental
comparison.

In some earlier works \cite{stiff1,stiff2} stiff polymers
were considered as perturbations about the straightline. This effectively
describes the tangent vector as varying on a tangent plane to the unit
sphere. This leads to the planar formula
\begin{equation}
P(\theta)= \frac{A}{LkT}\theta \exp{-\frac{A \theta^2 }{2LkT}},
\label{planar}
\end{equation}
for the distribution of final angles. Such a planar approach neglects 
the curved geometry of the sphere.
One also encounters a hybrid formula \cite{volo,mazur} 
\begin{equation}
P(\theta)= \frac{A}{LkT}\sin{\theta} \exp{-\frac{A \theta^2}{2LkT}},
\label{hybrid}
\end{equation}
in which one
computes $Q(\theta)$ using the planar approximation and includes the correct
curved measure $\sin{\theta}d\theta$ on the sphere.
The present treatment (unlike the earlier planar ones) deals with
geodesics on the sphere of tangents and thus takes into account the curvature
of the sphere {\it both} in the computation of $Q(\theta)$ and in the measure
$\sin{\theta}$. In the limit that the bending angle $\theta$  is small,
Eq.(\ref{Poftheta}) reduces to the earlier planar and hybrid formulae 
(\ref{planar},\ref{hybrid}). 
Our formula
for the bend angle distribution has a wider range of applicability
compared to the earlier ones: it is valid for  contour length 
comparable to the persistence length (stiff and the semiflexible regime) 
and also correctly describes the rare events involving 
large bend angles.

\section{Distribution of Bend Angles}
We have performed Monte-Carlo simulations using the 
Kratky-Porod model for comparison with the analytical form (\ref{Poftheta}). 
We randomly generated $10^6$ configurations of the 
polymer chain assuming it to be built of $N$ identical short straight 
segments each of length $l$.
Given one segment, the direction of the next segment was 
assumed to be uniformly distributed on a cone of 
semivertical angle $\Delta \theta=.1 $ radian around the preceding segment.
In the continuum limit when the number of segments
goes to infinity $N\ra \infty,l\ra 0,\Delta \theta\ra 0$
keeping $L=Nl$ and $L_p=\frac{2l}{(\Delta \theta)^2}$ fixed,
one recovers the wormlike chain model.
A frequency distribution of the number of configurations with 
angle $\theta$ between the initial and final tangent vectors was generated.

Figure 2 shows a comparison between our predicted analytical form, the planar
formula and
the results of computer simulation. In the stiff regime 
($\beta=L/L_p$ less than 1) eq. (\ref{Poftheta}) is virtually exact, 
in the semiflexible regime 
($\beta$ around $5$) it remains a reasonable approximation. 
Thus, in an experiment which takes repeated 
snapshots of the angle between the final and initial  
tangent vectors of a molecule, we expect
to find a distribution of angles given by Eq. (\ref{Poftheta}). This
is a prediction of the wormlike chain model and it goes beyond 
earlier studies based on classical elasticity\cite{mazur}.
\begin{figure}[h!t]
\includegraphics[width=8.0cm]{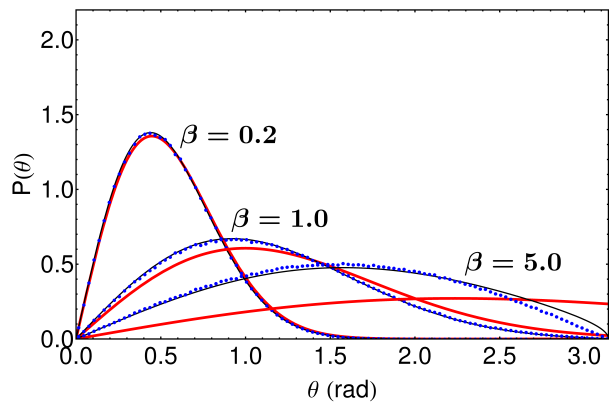}
\caption{(Color online) Comparison with simulations 
of $P(\theta)$ vs $\theta$ for a range of flexibilities. 
Plotted are the simulation data (blue dots), the planar formula (thick
red line) and our analytic formula Eq. (\ref{Poftheta})) 
(thin black line).
Notice that our closed form 
formula gives a reasonable fit to the 
simulation data even for $\beta=L/L_p$ as large as $5$.
}
\label{Pofthetafig1}
\end{figure}
\section{Moment Angle Relations}
The partition function $Q(\theta)$ can be converted into a free energy
by the formula (we drop $\theta$ independent terms as they are additive
constants in the free energy)
\begin{equation}
{\cal F}(\theta)= -kT \ln{Q(\theta)}=\frac{A \theta^2}{2L} -
\frac{kT}{2}\ln{\frac{\theta}{\sin{\theta}}}
\label{free}
\end{equation}
and used to calculate the bending moment 
$M$ needed
to bend the final tangent vector through an angle $\theta$.
The approximate closed form predicted by our theory is
\begin{equation}
M=\frac{A\theta}{L}-\frac{kT}{2}(\frac{1}{\theta}-\cot{\theta})
\label{momang}
\end{equation}
The first term in (\ref{momang}) represents the mechanical elastic
energy and the second (proportional to $kT$) is due to thermal
fluctuations.
\begin{figure}[h!t]
\includegraphics[width=8.0cm]{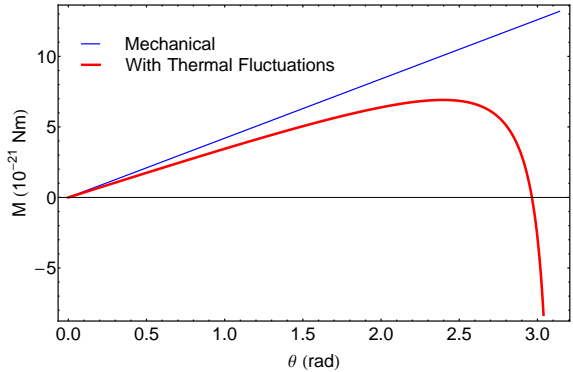}
\caption{(Color online) Bending moment vs bending angle theoretically
expected. The thin blue line shows the mechanical elastic response,
and the thick red line includes the effect of thermal fluctuations.
The moment is plotted in units of $10^{-21}$ N-m. 
The values of $A$ and $kT$ are chosen to reflect actin filaments under
physiological conditions.}

\label{momentangle}
\end{figure}

As Fig. 3 shows, the mechanical contribution is Hookean. Within 
a purely mechanical approach,
when the bending angle reaches $\pi$, the geodesic connecting $\hat{t}_i$ to
$\hat{t}_f$ is no longer the shortest one and
the configuration becomes unstable and buckles to a lower energy configuration.
This picture is drastically altered by thermal fluctuations. 
As shown by the red curve in Fig. 3, the buckling happens at an
angle smaller than $\pi$. The thermal fluctuations have the effect
of ``softening'' the linear response to external bending. 

\section{Conclusion}
We have analyzed the bending mechanics and fluctuations   
of semiflexible polymers. 
Our central goal in this paper has been to bridge the gap between 
mechanics and statistical mechanics by taking into consideration 
thermal fluctuation effects to quadratic order around mechanically
stable configurations.
To summarize, our main results are 1) an explicit 
formula for the bend angle distribution expected from the wormlike chain
model and 2) an explicit analytical formula for the moment
angle relation.

Our calculational technique assumes
that there is a unique shortest geodesic 
connecting $\hat{t}_i$ and $\hat{t}_f$.
This assumption breaks down at $\theta=\pi$. 
Our analytical formula
(\ref{Poftheta}) should only 
be used away from the point $\theta=\pi$
where there is a spurious divergence.
For a classical elastic rod, such as a poker or a ball point pen refill, 
one sees that if the bending angle exceeds $\pi$, the configuration
becomes unstable because the tangent vector no longer traces the shortest
geodesic. The poker then buckles into a new configuration with lower
energy. As shown in Fig. 3, thermal fluctuations cause the buckling to set in 
at bending angles considerably smaller than $\pi$.

For small bending angles $\theta$, one can Taylor expand the 
free energy (\ref{free}) and find that the effect of thermal fluctuations
is to effectively reduce the bending elastic constant $A$ 
by $kTL/6$. The fluctuations effectively soften 
the elastic response of the polymer.
For larger bending angles, one can no longer think of 
the fluctuations as simply renormalising the bending modulus $A$, 
since the form of the moment angle
relation is non Hookean.

While we have used actin as a typical example of a semiflexible
polymer, our study is also relevant to DNA and efforts to understand
the bending elasticity of small segments. Experiments and all atom
simulations \cite{mazur,cloutier,volo} 
are performed to understand the formation of ``kinks''
and cyclization of short and intermediate length DNA strands. 
Our present approach is analytical and computes
experimentally relevant quantities characterising the bending elasticity
of semiflexible polymers within the wormlike chain model. We expect our
results to interest researchers studying actin as well as DNA and
other biopolymers.   
\begin{acknowledgments} 
One of us (SS) acknowledges discussions with M. Santosh on all
atom simulations of DNA.
\end{acknowledgments}

\bibliographystyle{apsrev4-1}
\end{document}